\documentclass[twocolumn]{aastex61}
\usepackage[utf8]{inputenc}
\usepackage{amsmath}
\usepackage{amsfonts}
\usepackage{amssymb}
\usepackage{graphicx}
\usepackage{natbib}

\usepackage{times}
\hypersetup{linkcolor=blue,citecolor=blue,filecolor=blue,urlcolor=blue}

\newcommand{\RF}{ROUGH}
\newcommand{\muram}{MURaM}
\newcommand{\Al}{Alfv\'en}
\newcommand{\BP}{Bright Point}
\newcommand{\Bp}{Bright point}
\newcommand{\bp}{bright point}
\newcommand{\bhp}{bright-point}
\newcommand{\Bhp}{Bright-point}
\newcommand{\bz}{\ensuremath{B_z}}
\newcommand{\vz}{\ensuremath{V_z}}
\newcommand{\vx}{\ensuremath{V_x}}
\newcommand{\vy}{\ensuremath{V_y}}
\renewcommand{\vr}{\ensuremath{V_r}}
\renewcommand{\max}{\ensuremath{_\text{max}}}

\shorttitle{Bright-Point Motion at High Resolution}
\shortauthors{Van Kooten \& Cranmer}

\begin{document}

\title{Characterizing the Motion of Solar Magnetic Bright Points at High Resolution}
\author{Samuel J. Van Kooten}

\author{Steven R. Cranmer}
\affil{Department of Astrophysical and Planetary Sciences, University of Colorado, Boulder, Colorado, 80309, USA}

\correspondingauthor{Samuel Van Kooten}
\email{samuel.vankooten@colorado.edu}

\begin{abstract}
	
Magnetic \bp s in the solar photosphere, visible in both continuum and G-band images, indicate footpoints of kilogauss magnetic flux tubes extending to the corona.
The power spectrum of \bhp\ motion is thus also the power spectrum of \Al\ wave excitation, transporting energy up flux tubes into the corona.
This spectrum is a key input in coronal and heliospheric models.
We produce a power spectrum of \bhp\ motion using radiative magnetohydrodynamic simulations, exploiting spatial resolution higher than can be obtained in present-day observations, while using automated tracking to produce large data quantities.
We find slightly higher amounts of power at all frequencies compared to observation-based spectra, while confirming the spectrum shape of recent observations.
This also provides a prediction for observations of \bp s with DKIST, which will achieve similar resolution and high sensitivity.
We also find a granule size distribution in support of an observed two-population distribution, and we present results from tracking passive tracers which show a similar power spectrum to that of \bp s.
Finally, we introduce a simplified, laminar model of granulation, with which we explore the roles of turbulence and of the properties of the granulation pattern in determining \bhp\ motion.
	
\end{abstract}


\section{Introduction}
\label{sec:intro}
The surface of the Sun is the upper boundary of its convective zone, where convective cells called granules appear on spatial scales of 1~Mm and change on timescales of 10~minutes.
In the dark lanes between granules, bright, high-contrast structures are found, ranging from isolated points to extended structures wrapping around the granules.
These features, called ``G-band bright points,'' ``network bright points,'' ``magnetic bright points,'' ``magnetic bright features,'' ``inter-granular bright points,'' and ``filigree'', will be referred to simply as \textit{\bp s} in this paper.
Small, isolated \bp s are of order 0.1~Mm across and are visible for a few minutes at a time.
\Bp s are the footpoints of kilogauss-strength magnetic flux tubes extending into the corona.
In \bp s, the concentrated magnetic flux reduces gas pressure, lowering the optical-depth-unity surface below that of the surrounding areas by about 350~km \citep{Carlsson,Nordlund}.
This lower depth corresponds to a higher temperature (an $\sim800$ K difference), producing enhanced brightness relative to the surrounding downflow lanes.
This effect is visible in continuum imaging with sufficient resolution, but the contrast between \bp s and even the brightest granular areas is significantly enhanced in wavelengths such as the 430.5~nm G-band.
The G-band is dominated by lines from CH, a molecule whose abundance is significantly reduced at the optical bottom of \bp s through dissociation \citep{Steiner}.
This reduced CH opacity allows more light to shine through from the continuum below, producing significantly enhanced G-band contrast.

The dynamics of \bp s are important to study as a compelling potential source of coronal heating.
\Bp s move due to buffeting from the convective churning of the photosphere \citep{Berger}.
This motion transverse to the emerging flux tubes excites transverse magnetohydrodynamic (MHD) waves in the tubes.
These waves propagate up to the corona, where they are thought to turbulently dissipate, depositing heat.
Thus the power spectrum of \bhp\ motion is directly related to the power spectrum of these transverse waves.
This wave-heating power spectrum serves as an important boundary condition for models of transverse MHD wave propagation in the solar environment \citep[e.g.][]{Cranmer van Bal} and thus informs models of MHD-wave based coronal heating as well as models of the heliosphere.

A number of observations have been made of \bp s.
Approaches to understanding their motion include the use of passive tracers that follow the gradient of magnetic intensity images \citep{van Bal}, an approach which can handle extended structures; and the tracking of features identified by hand in individual frames \citep{Nisenson,Chitta}, which excels in directly tracking isolated \bp s of limited size.
More recent studies improve this latter approach with automatic identification of \bp s, producing much larger datasets \replaced{\citep{Utz,Keys,Yang 2014}}{\citep{Utz,Abramenko 2011,Keys,Yang 2014}}.

Observational determinations of the \bhp\ power spectrum are limited by the resolution and cadence of the observations.
Seeing conditions and pointing jitter can also limit the ability to track \bp s accurately over long durations.
In this paper, we use high-resolution simulations of the photosphere to produce a power spectrum of \bhp\ motion tracked at higher spatial resolution than has been done before.
This serves as a prediction of the observational \bhp\ power spectrum that the Daniel K. Inouye Solar Telescope \citep[DKIST;][]{Tritschler} will obtain with unparalleled spatial and temporal resolution.
We also introduce a simplified model of granulation, which allows investigation of the interplay between \bhp\ dynamics and the complex properties of the granulation pattern.
Section \ref{sec:muram} will describe our analysis of the high-resolution \muram\ simulation.
Section \ref{sec:rough} will introduce our simplified, laminar \RF\ model.
Section \ref{sec:results} will present our results, and Section \ref{sec:conclusion} will summarize our conclusions.

\section{\muram\ Simulations}
\label{sec:muram}
\subsection{Introduction}
\label{sec:muram_intro}

\begin{figure*}
	\centering
	\includegraphics[width=0.32\linewidth]{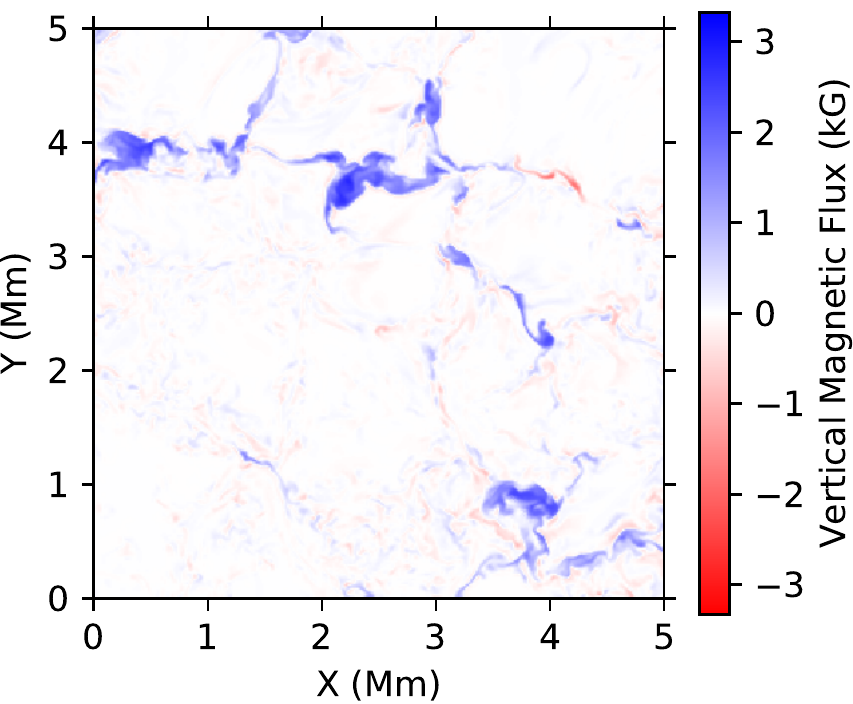}
	\includegraphics[width=0.32\linewidth]{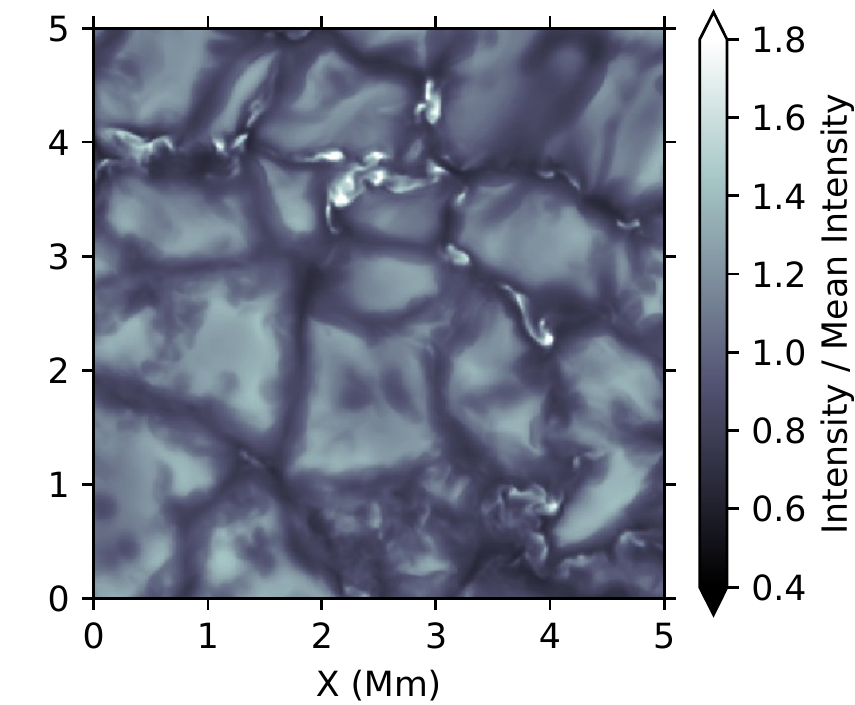}
	\includegraphics[width=0.32\linewidth]{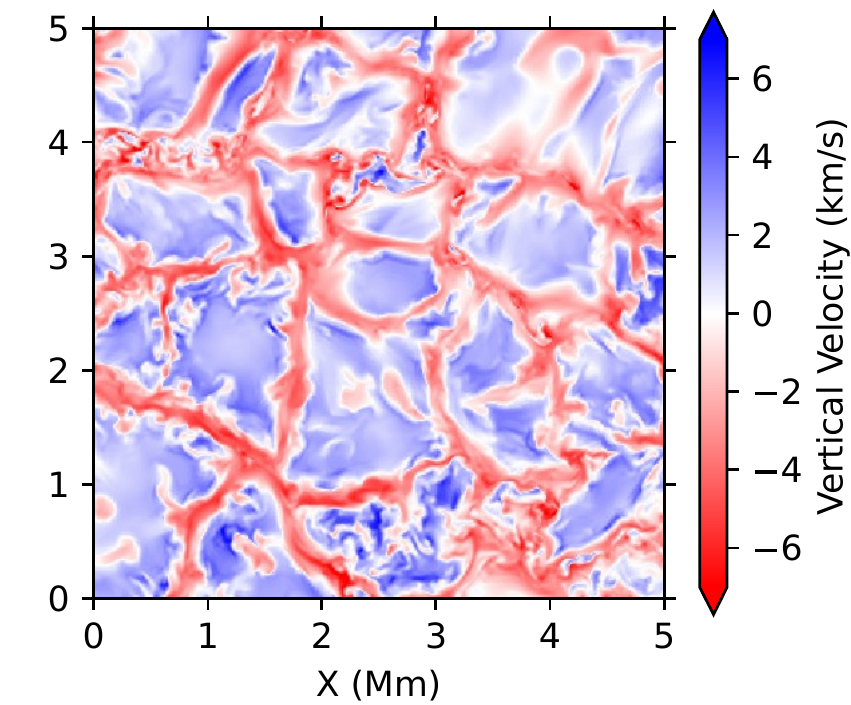}
	\caption{A common $5\times 5$~Mm portion of the \citet{Rempel} \muram\ simulation showing, from left to right, vertical magnetic flux, white-light intensity, and vertical plasma velocity at the beginning of the analyzed time range (time stamp 040000).  Over the full frame, peak values for $I/I_\text{mean}$ are near 2.75, peak values for \bz\ are near 3.3~kG, and peak values for \vz\ are near 9.9~km~s$^{-1}$ and -11.9~km~s$^{-1}$. (The full $24.5 \times 24.5$~Mm simulation is used in this paper.)}
	\label{fig:rempel_3_panel}
\end{figure*}

\citet{Rempel} uses \muram , software for performing radiative MHD simulations, to simulate the upper convective zone and photosphere.
The version of \muram\ used is extensively modified from the original of \citet{Vogler}.
In this paper we analyze the results of a run with a domain size of $24.576 \times 24.576 \times 7.680 \; \text{Mm}^3$, at a horizontal resolution of 16~km per pixel.
The data set covers 8280~s, or 2.3~hours, of time at a cadence of one snapshot every 20.7~seconds (for 401 total snapshots).
The simulation run begins with an imposed mean vertical magnetic field of 30 G (as discussed in Section 4.3 of \citet{Rempel}), but is otherwise identical to the ``O16bM'' described in the paper's Table 1.
The imposed flux is quickly swept into the downflow lanes, resulting in an enhancement of kilogauss features.
The top boundary is positioned 1.5~Mm above the $\tau = 1$ surface of the photosphere, allowing a convective region depth of 6.2~Mm.
From this simulation we take the $\tau=1$ surface as our solar-observation analogue.
Within this slice, we use the upward-pointing white light intensity and the three components of both plasma velocity and magnetic field.
A sample patch of the simulation is shown in Figure \ref{fig:rempel_3_panel}.

\subsection{Analysis Algorithms}
\label{sec:muram_algorithms}

From these \muram\ simulations, we extracted granule size distributions, the motion paths of passive tracers, and the motion paths of automatically-identified \bp s.
This section describes the algorithms used for collecting this information, while the data products themselves are analyzed in Section \ref{sec:results}.

For the identification of granules, observers such as \citet{Abramenko} use intensity from non-spectrographic data to segment granules.
With the \muram\ data, we segment granules based on vertical velocity instead, using just a positive-negative threshold.
While deviating from observational approaches, this produces very similar results with a simple, more direct technique.

We track passive, point-like test particles, called ``corks,'' as one way to analyze the plasma flows in the \muram\ data.
We begin with a uniformly-spaced grid of 4,900 corks.
(We find the power spectrum of cork motion to be very insensitive to the initial arrangement.)
Once initially placed and released, corks move according the horizontal portion of the velocity vector.
We linearly interpolate in time to a timestep of 4~s when propagating corks, to ensure corks in small, high-plasma velocity regions (e.g. turbulent whirls) do not maintain that high velocity after quickly exiting the region.
However, when computing power spectra, we sample the cork velocity every 20~s in order to avoid exceeding the Nyquist rate of the data.
With a timestep of 4~s, a pixel size of 16~km, and typical plasma velocities of $\sim2$~km~s$^{-1}$, a cork stays near its initial location during a timestep.
Cork locations are stored to floating-point precision, and the velocity field is interpolated spatially from the four nearest grid points.
Corks are allowed to travel through the periodic boundaries of the simulation.
Cork velocities used for computing power spectra are those read out of the interpolated velocity field at cork locations.

For tracking \bp s themselves, we use an algorithm that first identifies features within a given frame, following \citet{Feng}, and then links identified features between frames, following \citet{Yang 2014}.
The intra-frame algorithm keys in on the fact that \bp s are small and are characterized by high contrast (they are ``bright'' relative to their immediate surroundings, but not always in absolute terms---especially in continuum images).
In each step, we err toward preventing false positive detections at the cost of more false negatives, since there is such a wealth of \bp s available to be detected.

The single-frame stage begins by identifying ``seed'' pixels in a white-light intensity map.
A discrete analogue of the Laplacian is calculated by subtracting from each pixel's intensity the mean intensity of the eight surrounding pixels.
A local maximum in intensity will produce a significantly positive value, and this indicates a likely \bhp\ center.
We use a threshold of three standard deviations from the mean in the Laplacian to identify a set of likely centers.

\begin{figure}[t]
	\centering
	\includegraphics[width=\linewidth]{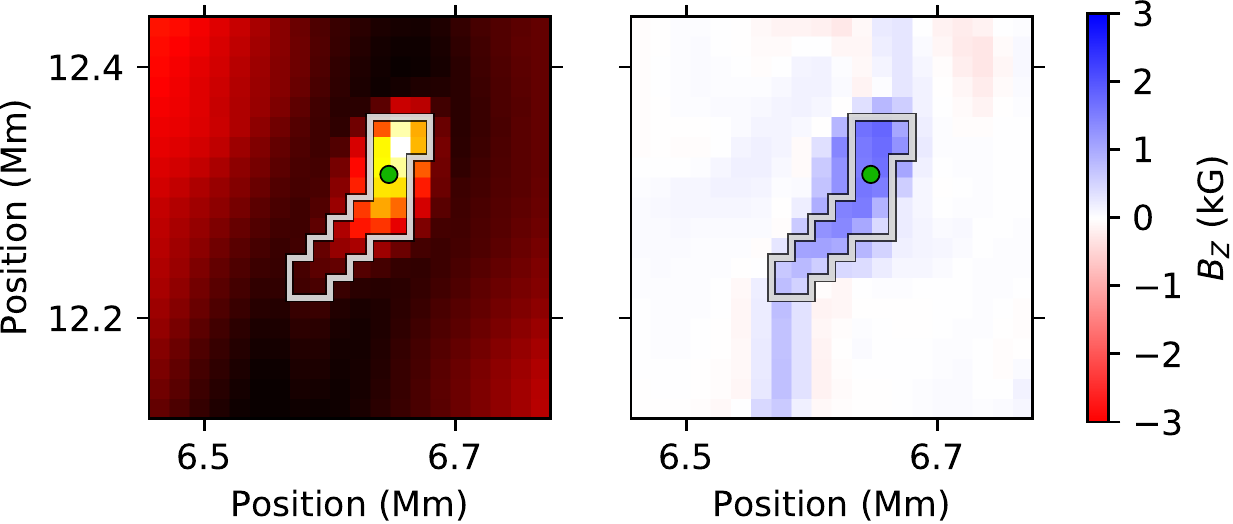}
	\caption{Intensity (left) and vertical magnetic field (right) for an example identified \bp. The gray line marks the borders of the identified region, while the green dot marks the intensity-weighted centroid. Identification is on the basis of intensity, and the vertical magnetic field shows the expected concentration.}
	\label{fig:bp_border}
\end{figure}

Next, the set of candidate pixels is dilated to include pixels that both exceed a lower threshold and neighbor an already-selected pixel (including neighbors along a diagonal).
We repeat this dilation process three times, and the threshold for inclusion in this stage is a positive Laplacian value.
This process expands the set of selected candidate pixels from the ``seed'' pixels, which are quite likely to be in \bp s, to a nearby cloud of pixels whose likelihood of being in a \bp\ is enhanced by virtue of being near a pixel already considered a candidate.
Since the structure of a \bp\ tends to be a decrease in intensity from a local maximum to low values at the edge, the use of the Laplacian $>$ 0 threshold aims to ensure that the dilation includes pixels within the \bp\ but stops expanding where it reaches the edge of the \bp.
At this point, each contiguous block of selected pixels can be identified and labeled as a unique candidate \bp.
An example of such a \bp\ is shown in Figure \ref{fig:bp_border}.

The next step is to remove as false positives local intensity maxima inside granules.
After the previous three rounds of dilation, these local maxima will produce identified features surrounded by further bright pixels.
True \bp s, however, tend to be mostly or entirely included within the identified set of pixels at this stage, so the identified feature will be surrounded by dark intergranular lane pixels.
(A \bhp\ diameter is of order 100~km, and a pixel in this simulation is 16~km across.
Three rounds of dilation allow a maximum expansion of $16\times3=48$~km or one \bhp\ radius, meaning a seed pixel at the center can be expanded to cover a typical \bp.
Often more than one seed pixel will be identified within a feature, allowing features on the large end of the spectrum to easily be captured.)
We compute the fraction of a feature's neighboring pixels which would be added to the feature if a fourth round of dilation occurred (still using the Laplacian $>$ 0 threshold).
We reject as false positives any features for which more than $20\%$ of these new neighboring pixels have positive Laplacians.

We apply two more criteria to ensure a high-quality selection of \bp s.
First, we reject features within four pixels of another feature.
This ensures the analysis is of \bp s that are cleanly and unambiguously separated from other \bp s.
We reject as short-lived transients features below 4 total pixels in area (36~km equivalent diameter).
We also reject features above 110 total pixels in area (189~km equivalent diameter) or above 20 pixels (320~km) across (as determined by finding the diagonal of the smallest rectangle to contain the feature), to ensure identified features are not so large or extended that their motion cannot be represented by the motion of their centroid.
(This necessarily eliminates from consideration the long chains of connected \bp s that can occur in regions of strong background magnetic flux; see Section \ref{sec:conclusion}.)

This completes the intra-frame identification step.
Each frame now has a set of uniquely-identified features, but these features must be linked in the temporal dimension for any analysis.
This linking step, following \citet{Yang 2014}, is short: if a feature in frame $n$ overlaps, by at least one pixel, the location of a feature in frame $n-1$, the two features are considered to be the same.
This is supported by observations \citep[e.g.][]{Chitta,Nisenson}, which find typical \bhp\ sizes of 100-150~km and typical velocity distributions with maxima around 7~km~s$^{-1}$. 
In the simulation's 20-second time steps, a \bp\ would move a maximum of 140~km (with most \bp s moving much less), ensuring that most \bp s will overlap with their prior position.
In situations where a feature in frame $n$ overlaps multiple features in frame $n-1$ (e.g. two \bp s merge), or where multiple \bp s in frame $n$ overlap one feature in frame $n-1$ (e.g. a \bp\ splits in two), linkage is ambiguous and is not attempted---instead, the feature(s) in frame $n$ are considered newly-found features.
Once linking is completed across all frames, we reject \bp s with lifetimes under 5 frames (1.6~minutes) as a further guard against transient features.

The result of this algorithm is a set of \bhp\ identification numbers, with each number corresponding to a set of coordinates within each frame of the \bp's lifespan.

\section{\RF\ Model}
\label{sec:rough}
\subsection{Introduction}
\label{sec:rough_intro}

In an attempt to accurately and efficiently capture the laminar portion of the solar granulation pattern, we developed a new Monte Carlo-based model called \RF\ (Random, Observationally-motivated, Unphysical, Granulation-based Heliophysics).
\RF\ simulations are highly phenomenological, and the properties of the simulation are drawn from observations of solar granulation.
The model aims to produce a reasonable representation of granulation at a single slice of constant height while offering very direct control over the properties of that granulation pattern.

In a \RF\ simulation, a set number of granule centers are randomly distributed over a region of space and throughout a given time span.
Granule centers are assigned a horizontal drift speed, drawn from a Gaussian distribution of zero mean and width 0.25~km~s$^{-1}$ \citep{Roudier 2012}, as well as a drift direction drawn from a uniform distribution.
Each center is also assigned a lifetime randomly drawn from the lifetime distribution of \citet{Del Moro}.
We use the exponential fit to lifetimes $>$ 5~minutes ($e^{-t/2.7 \; \text{min}}$, $t>5$~min), as this produces excellent agreement between simulated and observed granulation movies.
This produces a mean lifetime of 7.7~min.
At the end of a granule's lifetime, a granule can fade away or it can fragment.
\citet{Lemmerer} show that the majority of granules fragment into two or more child granules.
We give each simulated granule a 70\% chance of fragmenting into exactly two children.
The child granules each have new lifetimes drawn again from the \citet{Del Moro} distribution.

Each granule center is taken as a point source of radially-diverging, horizontal flow with a time-dependent amplitude of
\begin{equation}
	\vr (t) = V\max \; e^{- ( (t-t_\text{mid})/\delta)^4},
	\label{eq:V_r(t)}
\end{equation}
where $t_\text{mid}$ is the midpoint of the granule center's lifetime and $\delta$, set to a third of the granule's lifetime, determines the falloff rate with time.
$V\max$ is set to 5.1~km~s$^{-1}$ to match the integrated \RF\ power spectrum to the integrated \muram\ cork-tracking power spectrum.
The radial flow is given a spatial dependence of
\begin{equation}
	\vr (r, t) = \vr (t) \; \frac{r}{r_0} \; \frac{1}{(1 + (r/r_0)^a)^b},
	\label{eq:V_r(r, t)}
\end{equation}
where $r_0=500\;\text{km}$, $a=4$ and $b=1/a$ determine the shape of the function, and $r$ is the horizontal distance from a location to the center of the granule.
This functional form is chosen to match the observed granulation pattern \citep[see also][]{Simon}.
At each pixel in the simulation, the horizontal velocity is taken to be the velocity corresponding to the largest value of $\vr (r, t) / r$ produced at that pixel by any one granule center.
In other words, each granule center has a ``sphere of influence'' in which it is both the sole producer and the strongest possible producer of horizontal flow.
$\vr /r$ is used for this step rather than \vr\ itself to flatten the initial ramp-up from $\vr (r=0)=0$ to the function's maximum value.

\begin{figure}[t]
	\centering
	\includegraphics[width=\linewidth]{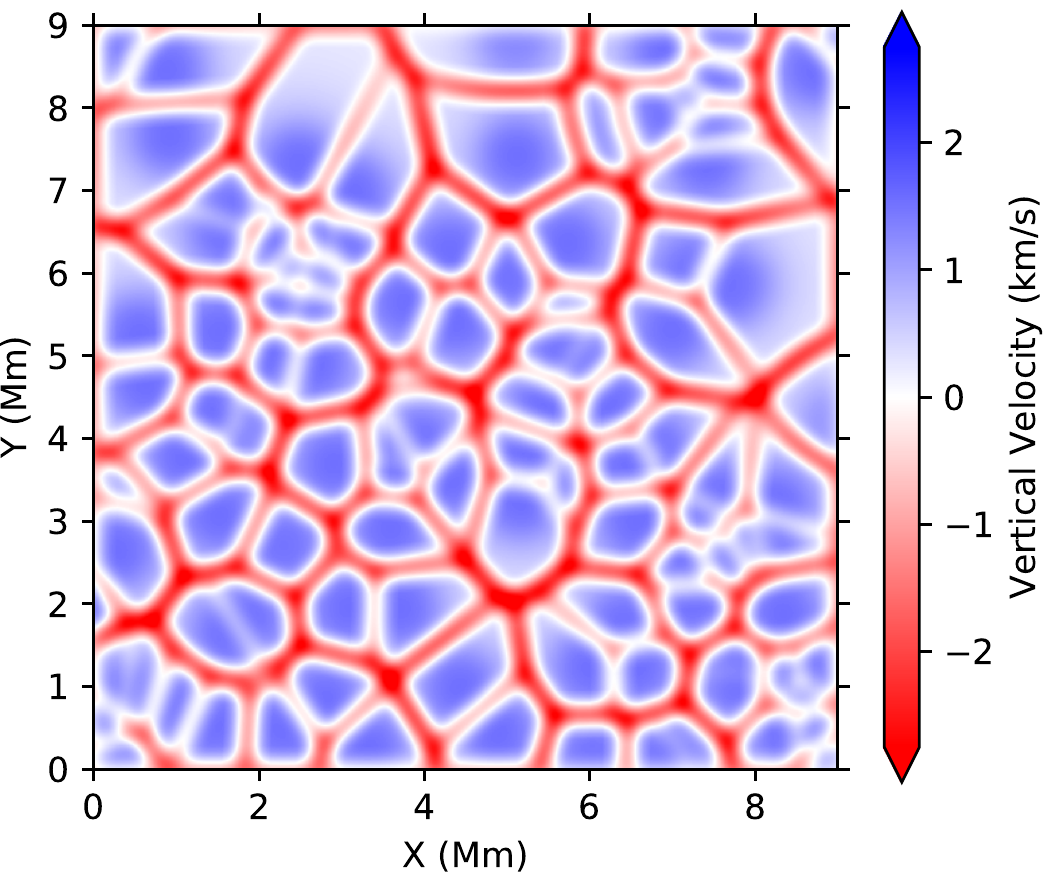}
	\includegraphics[width=\linewidth]{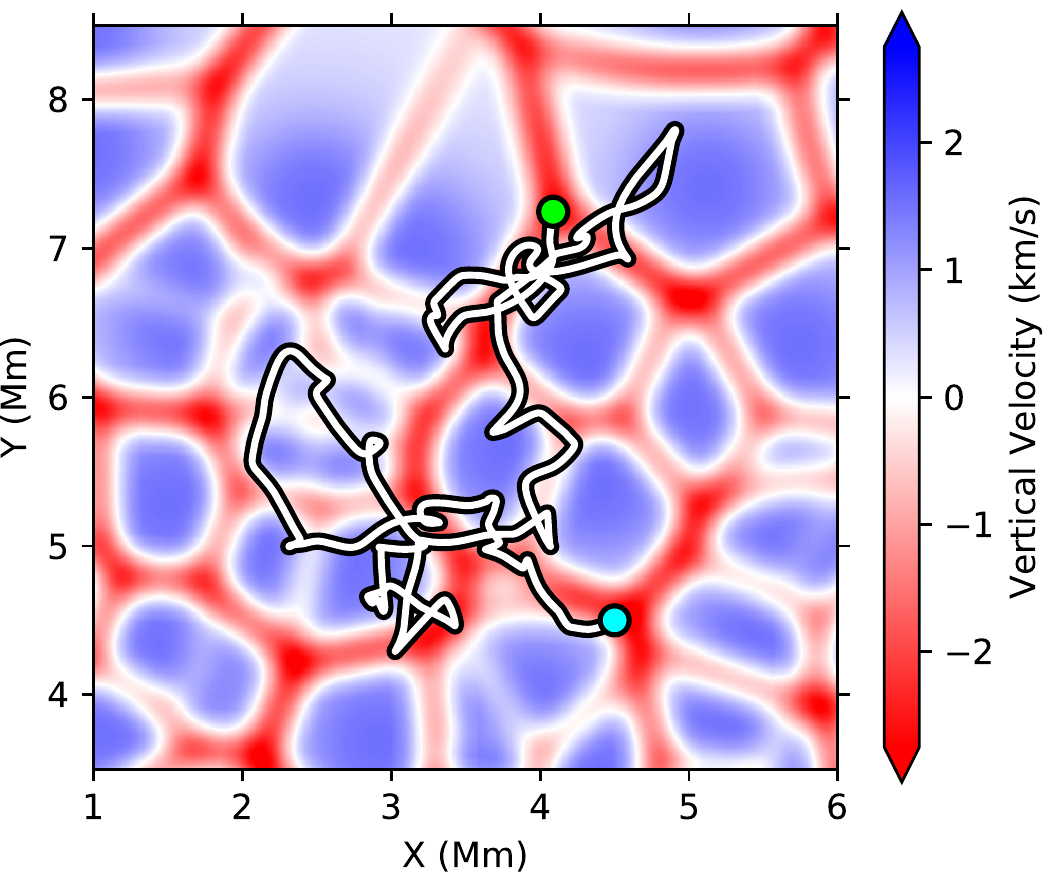}
	\caption{Sample granulation pattern produced by our \RF\ model. On the top, the full simulation domain. On the bottom, a magnified portion showing the motion track of a single cork over 166~minutes. The cork moved from the cyan dot on the \replaced{right}{bottom} to the green dot on the \replaced{left}{top}. The granulation pattern shown in both plots is that of the final frame.}
	\label{fig:Granulation_Pattern}
\end{figure}

At this point, the modeled granulation pattern will have discontinuous transitions at granule boundaries.
Gaussian smoothing with a $1/e$ full-width of 317~km is applied to the horizontal velocities to produce realistic inter-granular lanes of finite size.
An example of the resulting granulation pattern is shown in Figure \ref{fig:Granulation_Pattern}.

As a final step, vertical velocities are calculated by mass conservation in an assumed hydrostatic density profile, using
\begin{equation}
	\vz = H_\text{eff} \left(\frac{d\vx}{dx} + \frac{d\vy}{dy}\right)
	\label{eq:vertical velocity}
\end{equation}
\citep[see][]{Simon}, where the effective scale height $H_\text{eff}$, defined in terms of the pressure and vertical velocity scale heights by $1/H_\text{eff} = 1/H_P + 1/H_{\vz}$, is set to 75~km computed from values of $H_{\vz}$ $\approx$ 150~km \citep{Oba} and $H_P$ $\approx$ 150~km.

\RF\ cannot contain explicit \bp s, so instead we analyze the horizontal flow fields using passive tracers (``corks'').
A cork is placed initially at the center of the simulation domain.
At each time step, the horizontal velocity at the cork's location is calculated by a two-dimensional interpolation from the four nearest pixels, and the cork moves with that velocity until the next timestep.
The cork's location is not constrained to discrete pixel locations, but is calculated and saved to full numeric precision.
When calculating power spectra, we remove the initial portion of each cork's path, up until the cork first reaches a pixel with a downward vertical velocity.
This is more representative of actual \bp s, though we find that doing so produces little change in the final power spectrum.
An example cork motion track is shown in Figure \ref{fig:Granulation_Pattern}.

\begin{figure*}[t]
	% Stick this way up here for better pagination
	\centering
	\begin{tabular}{ccc}
	\includegraphics[trim={0 0 .9cm .8cm},clip,width=0.31\textwidth]{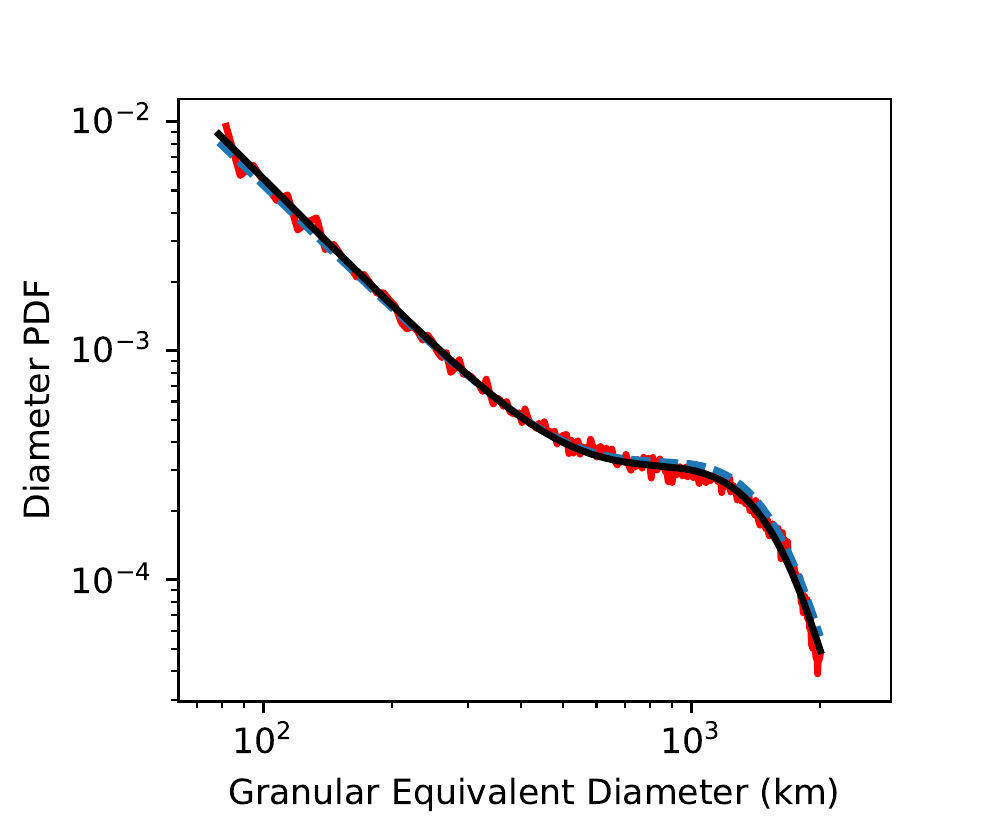}
	&
	\includegraphics[width=0.31\textwidth]{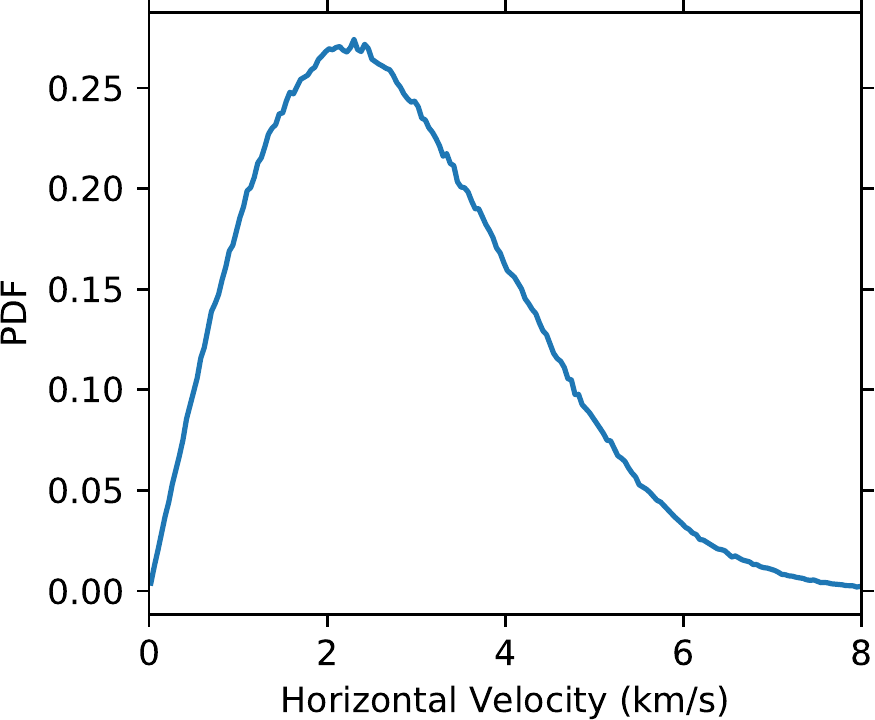}
	&
	\includegraphics[trim={0 0 0 .11cm},clip,width=0.31\textwidth]{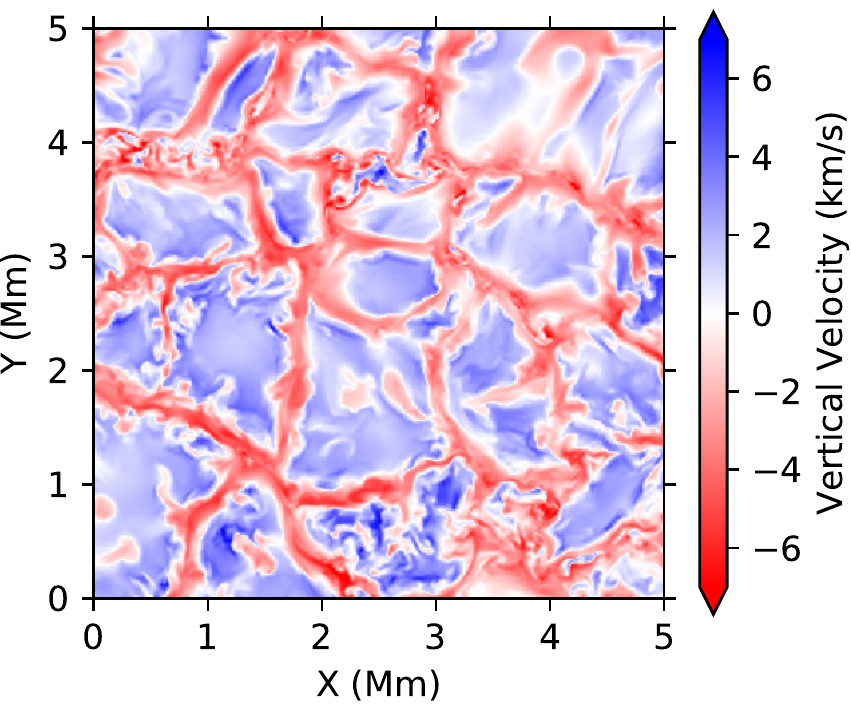}
	\\
	\includegraphics[trim={0 0 .9cm .8cm},clip,width=0.31\textwidth]{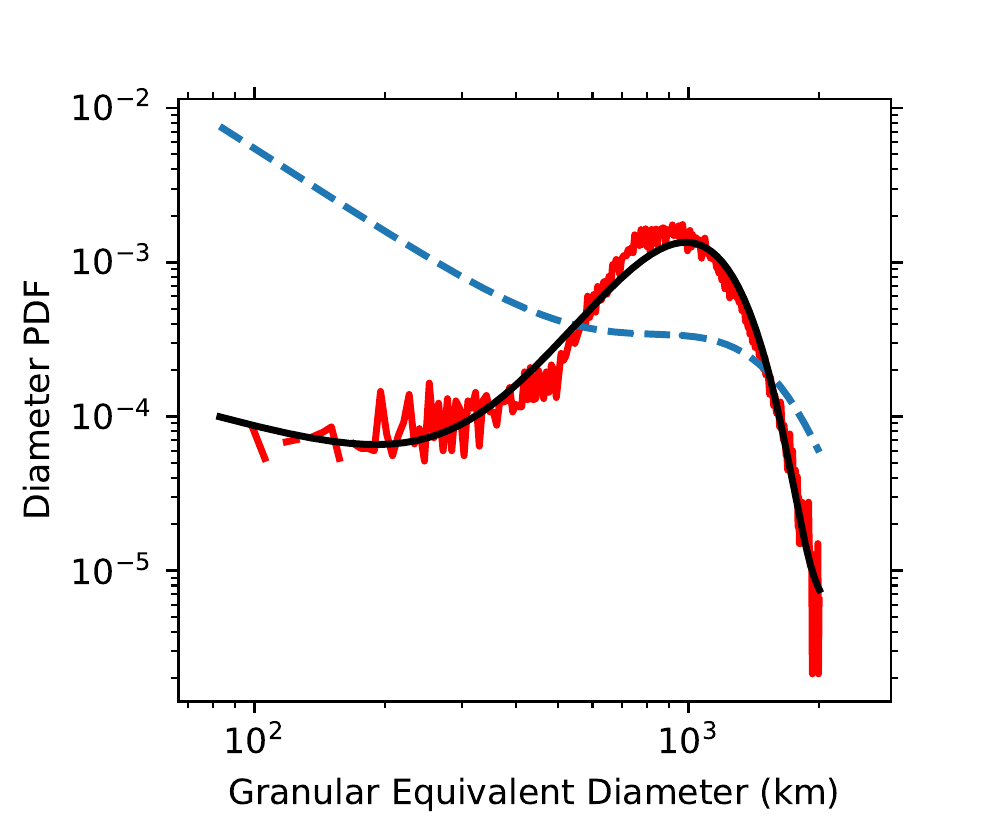}
	&
	\includegraphics[width=0.31\textwidth]{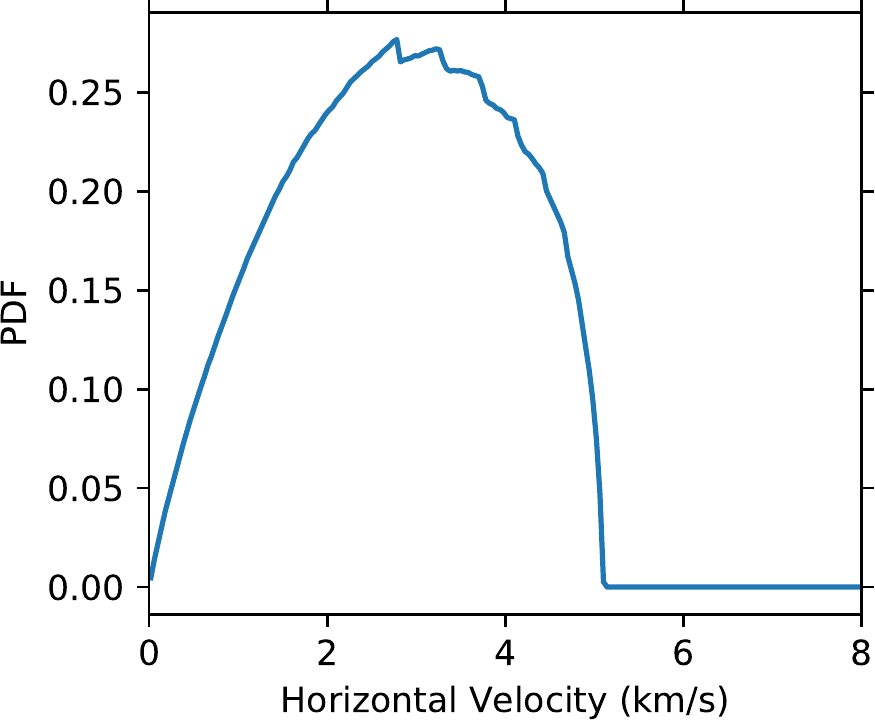}
	&
	\includegraphics[trim={0 0 .6cm 1.0cm},clip,width=0.31\textwidth]{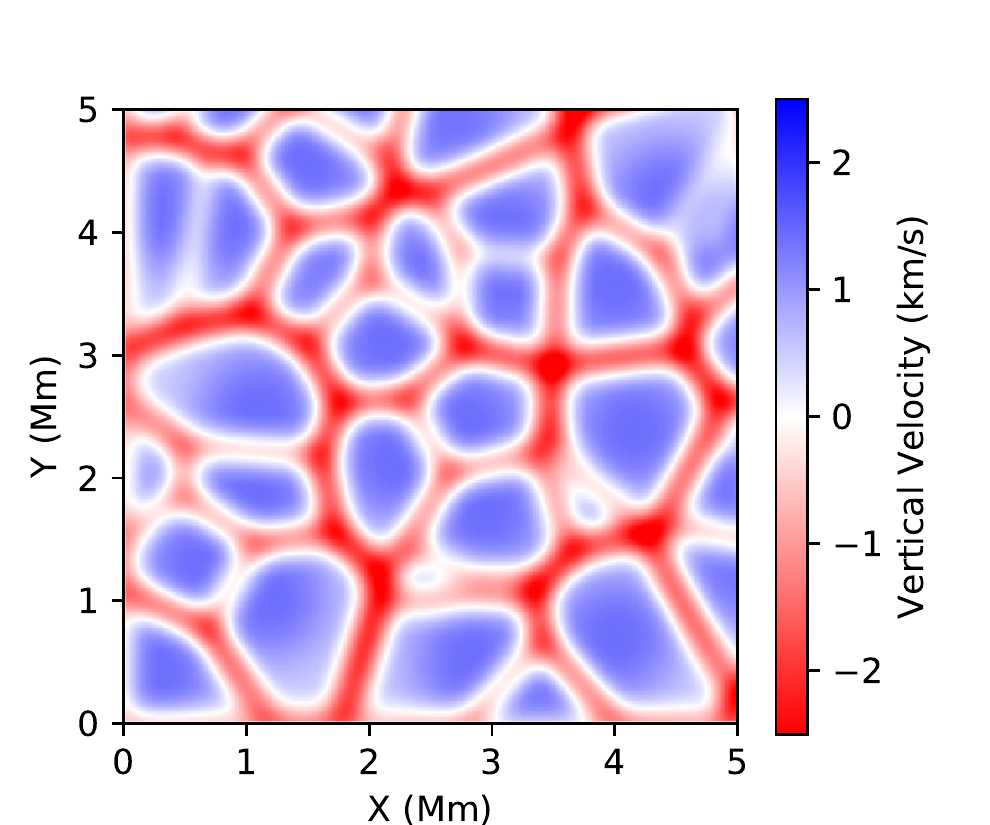}
	\end{tabular}
	\caption{Comparison of \muram\ (top) and \RF\ (bottom) granulation. On the left is the distribution of granule sizes. In red is the PDF of granule sizes. The dashed line is the fitted distribution of \citet{Abramenko} (re-normalized to this plotting range, and obscured by the other curves in the \muram\ panel), while the solid curve is the same model fitted to our histogram. In the center is the histogram of horizontal velocity magnitudes at all pixels in the simulation. On the right is a sample vertical velocity map. Sizes and horizontal velocities are drawn from the full 401 frames (2.3 hours) of \muram\ data, and from 1500 frames (2.5 hours) of \RF\ simulation.}
	\label{fig:six-pan}
\end{figure*}

The \RF\ model contains a number of adjustable parameters that control the granulation pattern, and varying those parameters yields physical insight.
We find that the cork power spectra are insensitive to the granule drift velocity distribution, consistent with the fact that granule drift is slow compared to granular flow velocities and to the timescale of change in the granulation pattern.
We find increased power in cork motion when the horizontal plasma velocities are increased, with the integrated power spectrum following a $\propto V_\text{max}^2$ scaling for moderate changes in velocity (reflecting the nature of kinetic energy).
We also find that total power is decreased when the granule emergence rate (per unit area, per unit time) is increased or when the granule scale size $r_0$ (Equation \ref{eq:V_r(r, t)}) is increased.
Increasing the emergence rate increases the number of granules and thus the fraction of area occupied by stationary inter-granular lanes.
Lanes are areas of converging horizontal flow, so a cork in a lane is in a relatively stable position and will experience less motion.
Further, having more granules causes each individual granule to be smaller.
When a new granule emerges and re-arranges the local intergranular lanes, then, the lanes will be shifted by a smaller distance and any displaced corks will be dragged along for a shorter distance before being in a stable location again.
An increase in the granule scale size pushes the maximum of the radial velocity function into the region affected by the Gaussian smoothing, thus extending the low-horizontal-velocity region from the center of granules to the edges.
This means that when the locations of the lanes changes, the granule-edge, horizontal flows that push corks along with the moving lane are weaker, and so corks move more slowly.

The analysis in this paper is of 3072 independent \RF\ simulation runs, with one cork per run.
We use a spatial resolution of 50~km/pixel, a time step of 2~seconds, a simulation duration of 167~minutes, and a granule emergence rate of $5.9 \times 10^{-9}$~km$^{-2}$~s$^{-1}$.

% This section header is on its own at the bottom of a column.
% Move it to the next column where its text starts.
\vfill\null

\section{Results}
\label{sec:results}

\subsection{Granulation}
\label{sec:results_granulation}
Here we present a comparison of the granulation patterns in \RF\ and \muram\ as a validation of the \RF\ approach.
In Figure \ref{fig:six-pan}, we compare the distribution of granule sizes between the two simulations and the observational values.
The observational values, from \citet{Abramenko}, draw on a field of view comparable to the \muram\ simulation domain and a diffraction-limited resolution of 77~km.
\citet{Abramenko} use intensity thresholding to separate granules from intergranular lanes.
As described in Section \ref{sec:muram_algorithms}, we instead use a $\vz=0$ threshold, which provides more straightforward granule segmentation.

Observed granule sizes are fit by the sum of a Gaussian distribution of large granules and a power law distribution of small granules \citep{Abramenko}, thought to reveal the presence of two distinct populations of granules \citep[as in][]{Hirzberger}.
We find that the \muram\ distribution reproduces very well the observational distribution.
However, the \RF\ distribution contains only the Gaussian portion.
This suggests that the small granules with a power-law size distribution have their origins in the turbulent phenomena not included in the \RF\ model, whereas the Gaussian-distributed large granules are produced by the large-scale, convective structures of the photosphere.

Figure \ref{fig:six-pan} also compares the distribution of horizontal velocities in the two simulations.
While the shape of the \RF\ distribution is set by the radial velocity functions (Equations \ref{eq:V_r(t)} \& \ref{eq:V_r(r, t)}), the horizontal scaling is set by the matching of integrated power spectra.
This produces good agreement between the two horizontal velocity distributions.
Notable, however, is the hard cutoff at $V\max$ in the \RF\ plot, whereas the \muram\ plot has a high-velocity tail due to turbulent regions.
Figure \ref{fig:six-pan} finally compares two granulation images, mapped in terms of vertical velocity, in which the qualitative agreement is good (aside from the presence of turbulence).

\subsection{\BP\ Statistics}
\label{sec:results_stats}
\begin{figure*}[t]
	\centering
	\includegraphics[width=0.32\linewidth]{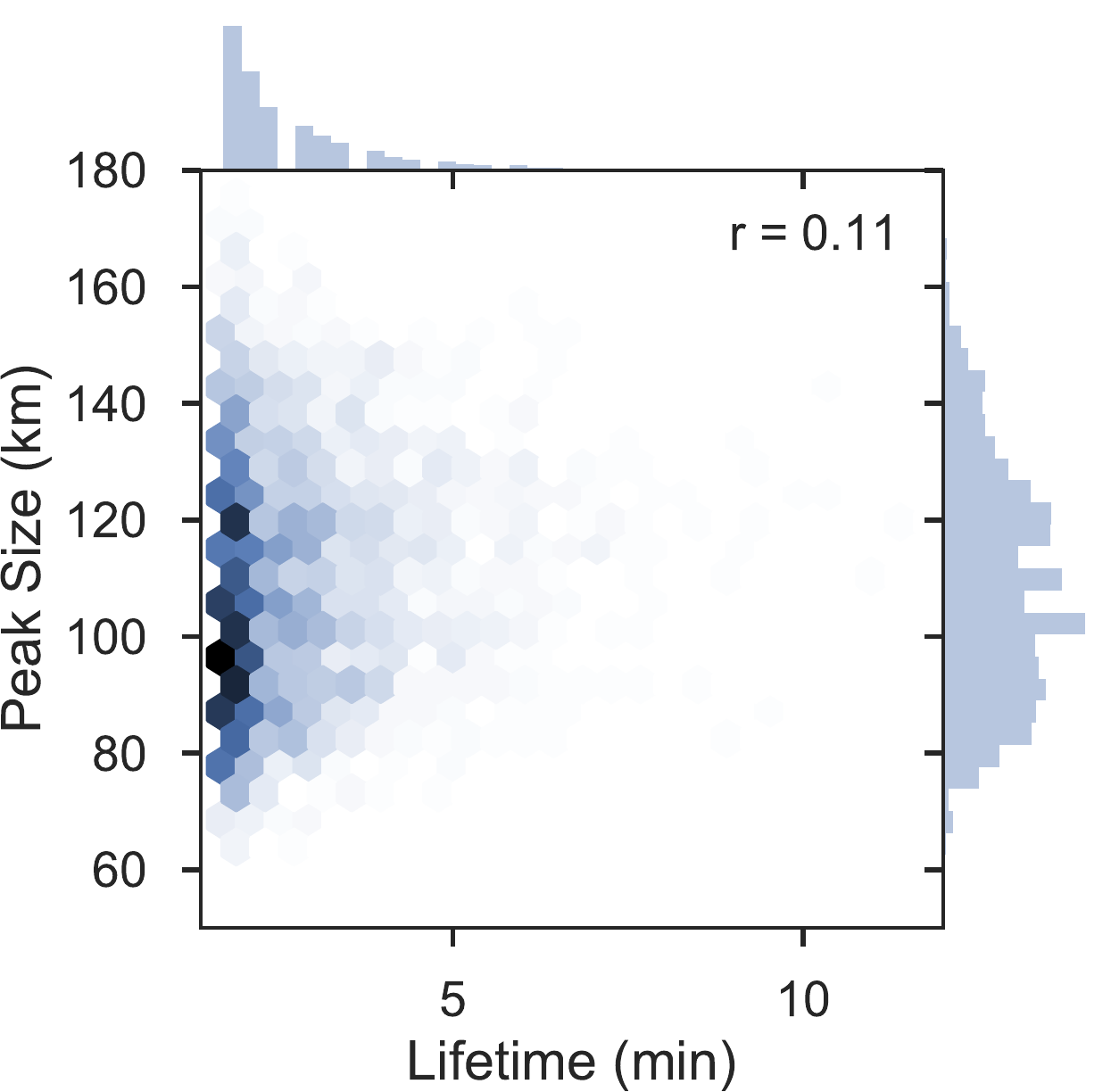}
	\includegraphics[width=0.32\linewidth]{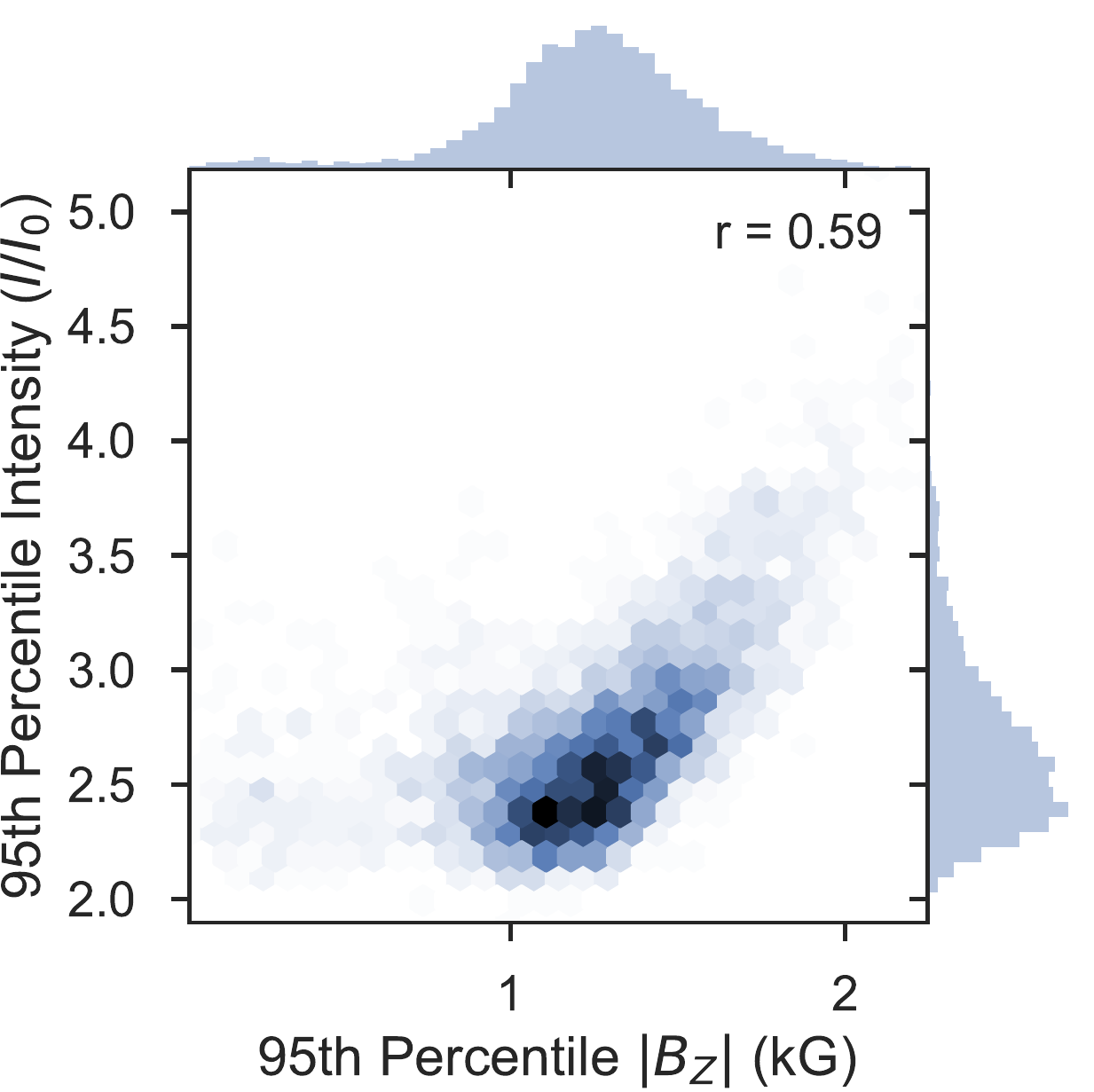}
	\includegraphics[width=0.32\linewidth]{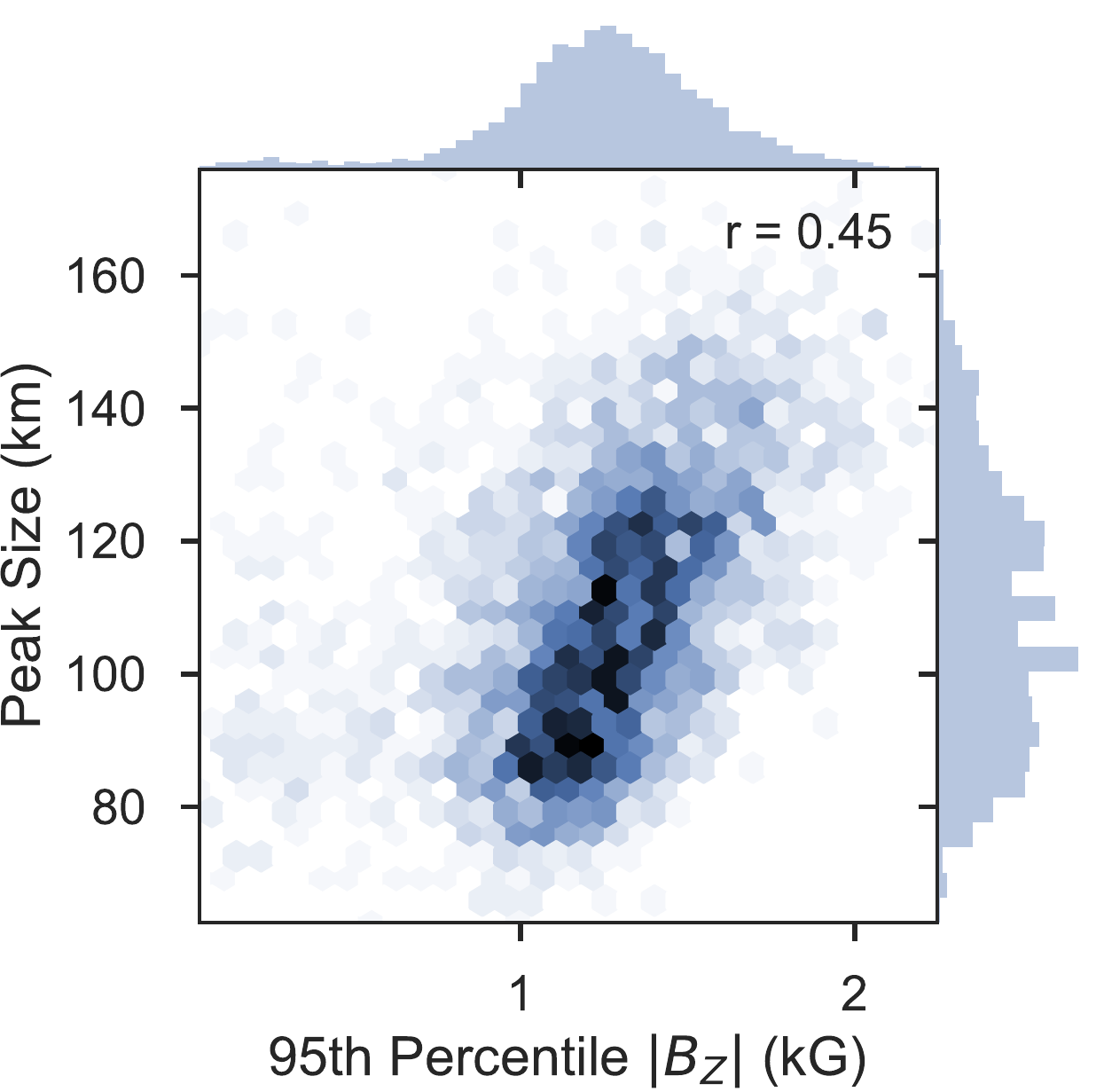}
	\caption{Distributions of \bhp\ statistics in the \muram\ data. ``Size'' is an equivalent diameter. The $95^\text{th}$ percentiles are computed from the values at all pixels identified as within a \bp . Marginalized distributions are shown along the edges of the plots. Shown for each distribution is the correlation coefficient; each has a two-sided p-value $<10^{-5}$.}
	\label{fig:bp_stats}
\end{figure*}

Applying the \bhp\ identification algorithm described in Section \ref{sec:muram_algorithms}, we find a total of 3,185 \bp s meeting all our size and lifetime thresholds in the \muram\ simulation data over the simulation time of 2.3 hours (401 snapshots).
Of these thresholds, the required lifetime of at least 1.6~minutes has the largest effect.
35 out of every 36 identified \bp s are removed by this criterion.
(See the steep lifetime distribution in Figure \ref{fig:bp_stats}.)
Of the remaining 3,185 \bp s, 9 were the result of mergers, 34 ended in a merger, 37 were the product of a fragmenting \bp, and 56 ended by fragmenting.

\begin{figure}[t]
	\centering
	\includegraphics[width=0.48\textwidth]{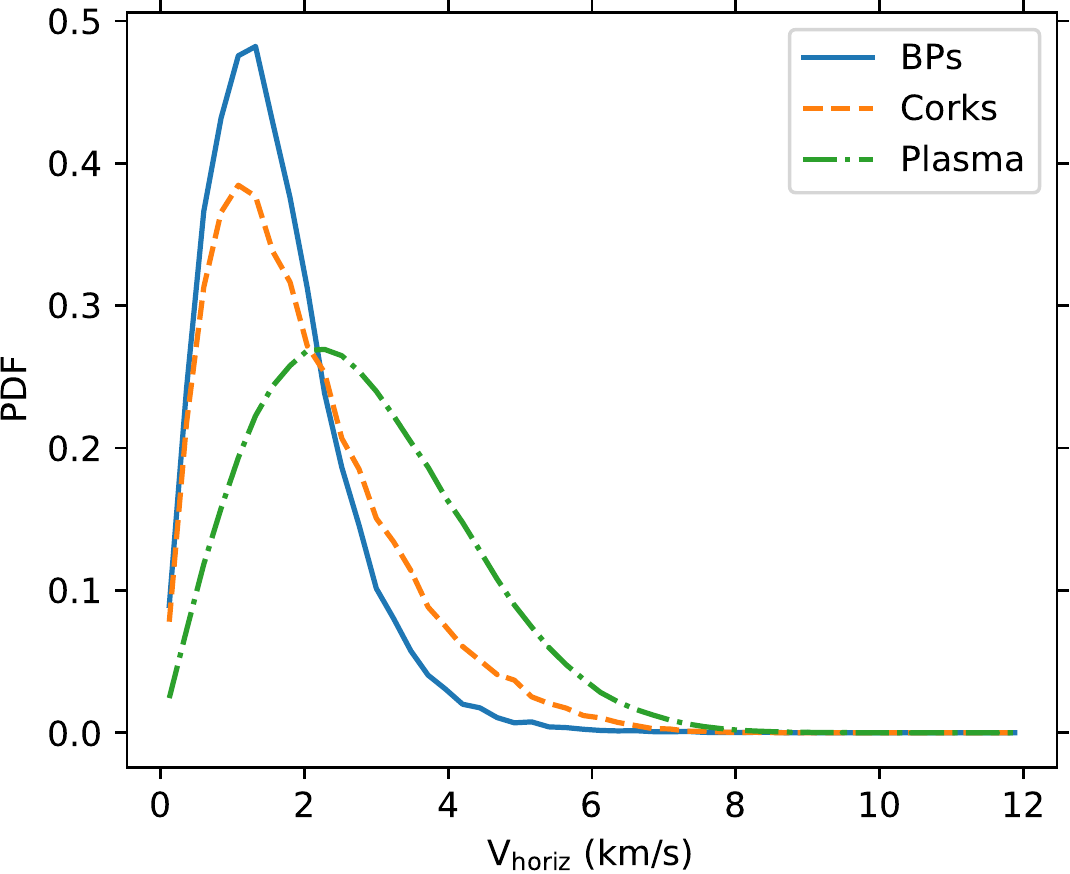}
	\caption{Horizontal velocity histogram for \bp s, corks, and plasma in the \muram\ data. Plasma velocity is sampled at every pixel in one frame, whereas the \bp and cork distributions use the full-length simulation. The largest value in each histogram is 8.97~km~s$^{-1}$ for \bp s, 10.8~km~s$^{-1}$ for corks, and 11.7~km~s$^{-1}$ for the plasma.}
	\label{fig:v_horiz_hist}
\end{figure}

In Figure \ref{fig:v_horiz_hist}, we show the distribution of observed horizontal velocities for \bp s, corks, and the plasma flow.
Cork velocities are lower than plasma velocities.
This is consistent with the fact that corks in a high-velocity region will travel at high velocity and quickly leave, while corks in a low-velocity region will linger; low-velocity regions are therefore preferentially sampled by corks.
\Bhp\ velocities may be susceptible to contributions from jitter in the centroid location caused by, e.g., \bhp\ shape changes or variation in the identified edge of the \bp .
We note here that \bhp\ velocities are typically lower than plasma or cork velocities, suggesting this is not a strong effect.
In the Appendix, we show more evidence that this effect appears to be insignificant in our data.

In Figure \ref{fig:bp_stats}, we show selected distributions of \bhp\ lifetimes, intensities, sizes, and vertical magnetic field strengths.
The distributions of \bhp\ size, vertical flux, and intensity all cluster near a central value (respectively, 109~km, 1.25~kG, and 2.7~$I_0$).
The \bhp\ lifetime, however, shows a sharp decay from the minimum lifetime threshold (1.6~minutes), with nearly all \bp s falling below a 5-minute lifetime.
The mean \bhp\ lifetime (for \bp s living longer than 1.6~min) is 2.7~min, while the maximum is 14.5~min.
This mean is comparable to the mean lifetime of $\sim$3~min for the longer-lived granule population in \citet{Del Moro}, indicating that \bp s evolve on the same timescale as the granulation pattern.
Our mean \bhp\ lifetime is notably shorter than the 9.33~minutes of \citet{Berger 1998}.
However, they take steps to track \bp s through merging and splitting as well as through short ($\sim 40$ s) disappearances of \bp s.
We do not attempt to match these capabilities, as doing so would complicate the calculation of power spectra.

Neither \bhp\ lifetime nor \bhp\ velocity show a strong correlation with any other measured value.
This may be expected for lifetime due to our inability to track \bp s through splitting and merging events, meaning that our reported lifetimes do not all correspond to the true lifetimes of \bp s, and due to the fact that \bp s themselves are only one stage in the constant evolution of magnetic flux.
The non-correlation may also be expected for velocity, since \bhp\ motion is controlled by external forces.
However, strong correlations emerge between \bhp\ intensity, size, and magnetic field strength.
We use for each \bp\ the maximum size and the 95$^\text{th}$ percentile pixel value value for intensity and field strength in order to capture the value when the \bp\ is most fully emerged and vertical but to avoid outlier values.
We find a correlation coefficient of 0.59 between intensity and $|\bz|$, 0.44 between size and $|\bz|$, and 0.62 between size and intensity.
These correlations are all statistically significant and reasonably suggest that a stronger flux concentration will produce a larger, brighter \bp\ \citep[a conclusion supported observationally, e.g.][]{Ji}.

The correlations we find between \bhp\ lifetime and both diameter and velocity differ from those found in \citet{Yang 2014}.
These Hinode/SOT observations yielded a correlation coefficient between lifetime and size to be +0.83 (contrasted with our value of +0.01 for mean values and +0.1 for peak values) and between lifetime and velocity to be $-$0.49 (contrasted with our $-$0.09 for mean values and +0.14 for peak values).
The large differences here are likely due to a difference in approach.
\Bhp\ lifetimes are discrete multiples of the imaging cadence, and for each lifetime value, \citet{Yang 2014} calculate a mean value of diameter and velocity among all \bp s of that lifetime and use these mean values to perform a linear regression.
In contrast, we use our full dataset with all its variation.
Using average values collapses the data to a mean trend line, which risks artificially enhancing the resulting correlation coefficient.
We believe that using the full set of data more accurately represents the level of variation in the data (and, therefore, how well one measurable can predict another).

\subsection{Power Spectra}
\label{sec:results_spectra}

Following the literature \citep{van Bal,Cranmer van Bal,Chitta}, we compute power spectra from velocity sequences via the Wiener-Khinchin theorem, which provides a power spectrum as the Fourier transform of the autocorrelation of a data sequence.
While perhaps a roundabout approach\footnote{In fact, this is the reverse of the usual application of the Wiener-Khinchin theorem, which is often used as an efficient way to calculate an autocorrelation by way of a Fourier transform, an absolute-square, and an inverse Fourier transform.}, this allows individual \bhp\ velocity sequences to be more straightforwardly averaged together.
Since velocity sequences of different lengths are produced, their direct Fourier transforms (and those transforms' absolute-square) have different ranges and spacings in frequency, making a more direct averaging non-trivial.
But as long as different velocity sequences have the same temporal spacing, their autocorrelations will have the same spacing.
This allows averaging to be performed on the autocorrelations (analytically equivalent to averaging the final power spectra), which is computationally simpler.

For the \muram\ \bp s, we calculate power spectra of the finite-difference velocities of the intensity-weighted centroids of \bp s.
Our observational comparison, \citet{Chitta}, present their power spectrum and autocorrelation function as fitted, analytical functions.
In this section, we present the power spectrum produced by discretely sampling the \citet{Chitta} autocorrelation function at its observational cadence of 5 s out to $\pm$ 105 s (the range used for their fit to the autocorrelation) and Fourier transforming, to provide a more direct comparison to our own power spectra.

\begin{figure}[t]
	\centering
	\includegraphics[width=\linewidth]{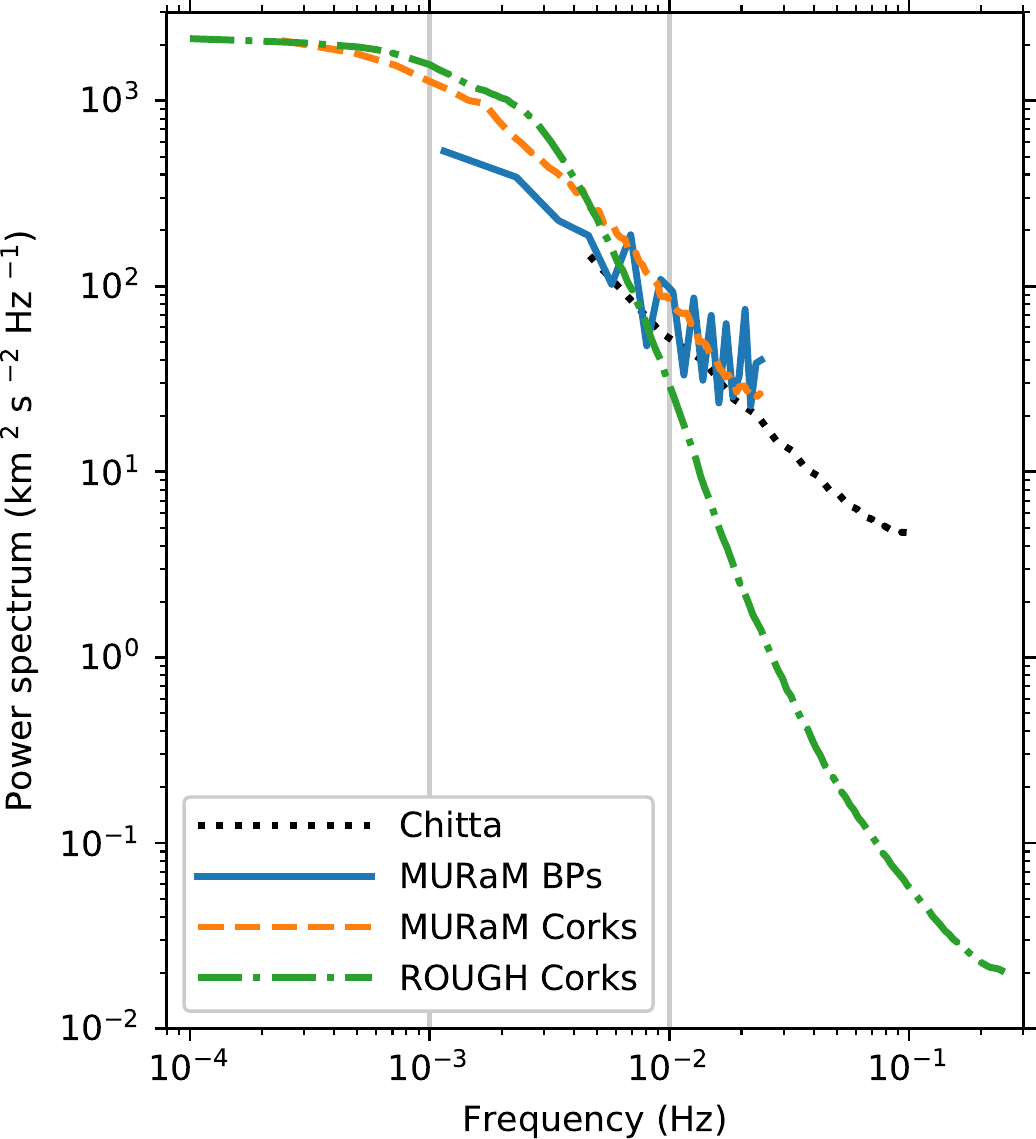}
	\caption{All power spectra mentioned in this paper. The \muram\ cork curves are those of Section \ref{sec:muram_algorithms}. The \muram\ curve is from our \bp\ tracking approach described in Section \ref{sec:muram_algorithms}. The Chitta curve is from the observations of \citet{Chitta}. 
	The \RF\ curve comes from our simulations, described in Section \ref{sec:rough_intro}. Gray vertical lines mark the frequencies of the slope measurements in Table \ref{tab:integrated_power}.}
	\label{fig:all_spectra}
\end{figure}

Power spectra for observations, \muram\ corks, \muram\ \bp s, and \RF\ corks are shown in Figure \ref{fig:all_spectra}.
\RF\ is configured so that the integrated power matches the integrated \muram\ cork power.

At low frequencies, the \RF\ spectrum is similar to the \muram\ cork spectrum, whereas at high frequencies it drops off precipitously.
This reflects the similarity between the two simulations of the large-scale granulation patterns, which cause low-frequency movement, and the absence in \RF\ of small-scale turbulence \citep[e.g. vortex flows, see][]{Giagkiozis}, which drive high-frequency motion.
This comparison also suggests that these two phenomena are the dominant drivers of motion in their respective frequency regimes.

The \muram\ cork spectrum shows more low-frequency power than the \muram\ \bp s, suggesting that perhaps corks are more free to drift in steady flows than are \bp s, but the two spectra converge at high frequencies, suggesting that they respond similarly to rapid variation in plasma flow.
Finally, both \muram\ analyses show a consistent enhancement in power \deleted{(by $\sim$50\%) }over the Chitta data for the frequencies of overlap\added{, quantified later in this section}.

The integral of the power spectrum is equal to the variance of the velocity data (via Parseval's theorem).
These integrated values are shown in Table \ref{tab:integrated_power}.
\added{We report both the integral of each spectrum over its full extent and the integral over the frequency range covered by all four spectra (0.005 Hz $\le$ f $\le$ 0.02 Hz).
The former represents a spectrum's lower bound on the ``true'' total power, unconstrained by limited frequency coverage, while the latter allows spectra to be compared over a comparable domain.}
\replaced{This reveals}{These numbers reveal} more motion and variation for \muram\ corks than for \bp s, and as noted above, \replaced{more motion for both \muram\ analyses than for the observed \bp s}{we find a 91\% increase in total power measured in the frequencies of overlap for \muram\ \bp s versus observed \bp s}.
This increase in motion, in both the power spectra and the integrated spectra, might be explained in part by the calibration offset applied by \citet{Chitta}, a step which could tend to damp out some real motion in addition to spurious motion and which is not needed in simulation analysis.
It might also suggest there is, in fact, more power to be detected with higher-quality observations.
Improvements in spatial resolution and sensitivity will make \bhp\ detection and centroid identification more reliable and make smaller \bp s visible, both enabling more robust statistics.
Verification of the presence of unmeasured power must wait until DKIST comes online, currently scheduled for 2019.
The Visible Broadband Imager on DKIST will have a diffraction-limited, G-band resolution of $\sim$16~km (comparable to these \muram\ simulations) and a 3~second cadence \citep{Elmore}, improving on both the $\sim$160~km resolution of the Hinode SOT observations of \citet{Yang 2014,Yang 2015} and the $\sim$100~km resolution and 5~second cadence of the SST observations of \citet{Chitta}.  

Both the \muram\ and \citet{Chitta} power spectra shown in Figure \ref{fig:all_spectra} indicate power-law behavior with slopes of order $f^{-1}$ between $10^{-3}$ and $10^{-2}$~Hz (see Table \ref{tab:integrated_power}).
This is a much flatter spectrum than the presumed exponential drop-off associated with random-walk-like motions like those found from earlier measurements of bright points \citep[e.g.,][]{van Bal,Cranmer van Bal}.
However, whether this points to the presence of an active turbulent cascade \citep{Petrovay} in the photosphere is still not known.
Reduced MHD turbulence simulations for solar flux tubes \citep{van Bal 2011,Woolsey} showed that low-frequency random-walk motions rapidly produce a steep power-law spectrum ($P \sim f^{-4.5}$) in the photosphere.
This may be modified for the flatter bright-point driving spectrum found in this work.
There is a long history of interplanetary $f^{-1}$ fluctuations being interpreted as fossil remnants of oscillations that map down to the solar surface \citep{Matthaeus,Velli,Dmitruk,Verdini}.
However, those fluctuations occur at even lower frequencies than those studied in this paper (e.g., $f < 10^{-4}$~Hz) and their physical origin is still being debated.

\begin{table*}[t]
	\centering
	\caption{Properties of the spectra in Figure \ref{fig:all_spectra}.}
	\label{tab:integrated_power}
	\begin{tabular}{lDDll}
		\hline \hline
		Spectrum & \multicolumn{4}{c}{Integrated Power (km$^2$ s$^{-2}$)} & \multicolumn{2}{c}{Slope at} \\
		& \multicolumn{2}{c}{over all frequencies} &  \multicolumn{2}{c}{0.005 Hz $\le$ f $\le$ 0.02 Hz} & 0.001 Hz & 0.01 Hz \\
		\hline
		\decimals
		Chitta & 1.87 & 0.56 & \nodata & -1.25 \\
		\muram\ BPs & 3.46  & 1.07 & \nodata & -1.27 \\
		\muram\ Corks & 5.42\tablenotemark{a}  & 1.21 & -0.37 & -1.59 \\
		\RF\ Corks & 5.42\tablenotemark{a}  & 0.57 & -0.45 & -3.45\\
	\end{tabular}
	\tablenotetext{a}{\RF 's $V\max$ value is set so its corks' integrated spectrum matches that of the \muram\ corks.}
\end{table*}

\section{Discussion and Conclusions}
\label{sec:conclusion}

The goal of this paper was to better understand the horizontal motions of photospheric intergranular \bp s.
We characterized this motion using simulations that exceed present observational capabilities, and we investigated which aspects of the granular flows drive this motion.

Using high-resolution simulations, we found an increase of \replaced{$\sim$50\%}{91\%} over observations in the power spectrum of \bhp\ motion at all frequencies of overlap.
This suggests the possibility that the limitations of current observations under-represent the motion of \bp s---an idea to be tested by DKIST observations.
The power spectrum of \muram\ \bp s is only negligibly affected when blurring the simulation output to an observational resolution.
However, the expected increase in signal-to-noise ratio for DKIST observations also plays a significant role in improving \bhp\ tracking reliability.

We compared the power spectra of passive tracers in \muram\ flows and \RF 's simplified, turbulence-free flows.
This comparison suggested that low-frequency motion of passive tracers (and thus \bp s) is driven by long-term granular evolution, while high-frequency motion is due to turbulent effects.

We found that the power spectra of \bhp\ motion and passive-tracer motion are very similar at high frequencies, but passive tracers display about 2.5 times more power at low frequencies.
One possible explanation for this is that the passive tracers move according to the plasma flow field at the $\tau = 1$ surface only, whereas \bp s are pushed at multiple depths along the flux tube.
Any depth dependence to the horizontal plasma velocity will result in competing forces on the flux tube, producing muted \bhp\ motion.
That this difference is seen only at low frequencies may suggest that rapid changes to the flow field (producing high-frequency motion) are either coherent across the relevant depths or are sufficiently strong at one depth to dominate over the flows at other depths, whereas long-term flows have more depth dependence and so produce net smaller effects on \bp s.

We found that \muram\ effectively reproduces the granule size distribution of the observations in \citet{Abramenko}.
However, of this two-component distribution, \RF\ reproduces only one component---the Gaussian-distributed, large granules.
We proposed that the two components to the granule size distribution represent the large convective cells and the small, turbulent features.
Separately, \citet{Del Moro} finds that granule lifetimes are best fit by a two-exponential model, finding a different fit for the $> 5$~minute and $< 5$~minute lifetimes.
We speculate that these two fits match the two classes in the granule size distribution---the small, turbulent features correspond to the short-lived features, and the large-scale convective features correspond to the long-lived features.
Indeed, \citet{Del Moro} finds a correlation between granule lifetime and size that is positive for short-lived granules, but which flattens for lifetimes over $\sim 8$~minutes.

We found that \bhp\ size, brightness, and magnetic flux are all positively correlated (with $r \sim 0.5$).
This can be made sense of, remembering that \bp s are bright because a reduction in gas density lowers the $\tau =1$ surface to hotter, brighter regions.
A larger-diameter flux tube will tend to allow observations deeper into the tube before bending cuts off the line of sight, while stronger magnetic fields will produce lower gas pressure and density in the flux tube that allows a clearer view deeper into the tube.

It is worthwhile to use the results found above to investigate some properties of \Al\ fluctuations that propagate up into the corona and give rise to dissipative heating.
Following \citet{Cranmer van Bal}, it is possible to estimate the net upward energy flux $F$ of \Al\ waves given the properties at the photosphere,
\begin{equation}
  F \, = \, \rho \langle v_{\perp}^2 \rangle V_{\rm A}
  \left( \frac{1 - {\cal R}}{1 + {\cal R}} \right) \,\,\, ,
  \label{eq:flux}
\end{equation}
where we assume a photospheric density $\rho = 2 \times 10^{-7}$ g~cm$^{-3}$ and typical bright point field strength $B = 1500$ G in order to estimate the \Al\ speed $V_{\rm A} = B/\sqrt{4\pi\rho}$, and the velocity variance $\langle v_{\perp}^2 \rangle$ is the integrated power given in Table \ref{tab:integrated_power}.
The above expression also depends on the reflection coefficient for wave energy ${\cal R}$, which \citet{Cranmer van Bal} found to be approximately 0.90 below the transition region.

Note that Equation (\ref{eq:flux}) gives the energy flux {\em inside} a vertically oriented bright-point flux tube.
However, in the corona, these flux tubes are expected to broaden out to eventually fill the entire volume.
Thus, a more relevant energy flux for coronal heating is the surface-averaged value $\langle F \rangle \approx f_{\ast} F$, where $f_{\ast}$ is the photospheric filling factor of bright points\added{, which we treat as a simple, unweighted area fraction}.
In the \muram\ simulation analyzed above, $f_{\ast} = 0.00474$ for all identified bright points, including those with lifetimes below 1.6~minutes that did not yield useful velocity information.
Combining the above quantities, we find $\langle F \rangle \approx 860$ W~m$^{-2}$ for the \citet{Chitta} integrated power, 1590 W~m$^{-2}$ for the power in \muram\ bright points, and 2490 W~m$^{-2}$ for the power in \muram\ cork motions.
These values are well above the classical \citet{Withbroe} empirical requirements for coronal heating in the quiet Sun (300 W~m$^{-2}$) and coronal holes (800 W~m$^{-2}$), indicating that only a fraction of the wave energy flux needs to be dissipated as heat in these regions.

One aspect not considered in this work is the impact of long chains of connected \bp s seen in regions of strong background magnetic flux \citep{Dunn,Almeida}.
These features were considered by \citet{van Bal}, but they generally require a different approach than that used in this paper for small, isolated \bp s.
Differences may be expected between the dynamics of the isolated and the extended types of \bp s, due to both interactions between flux elements in extended structures, and the observed trend for \bp s to move more slowly in regions of increased magnetic flux \citep{Ji,Yang 2016}.

While the transverse \bhp\ motion studied in this paper is useful, it does not tell the complete story.
MHD waves can also be excited both inside and outside the \bhp\ flux tubes by compressive and longitudinal fluctuations.
This fact will become more prominent when DKIST turns traditionally-point-like \bp s into resolved structures in which the motions necessary to excite these additional modes will become observable.
In fact, as discussed in the Appendix, the ability to track higher-order motion may become necessary.
We plan to study these wave modes in future work.
Beyond MHD waves, \bhp\ motion can also be analyzed in bulk, looking at the diffusive statistics of the \bp s \citep[e.g.][]{Abramenko 2011,Ji,Agrawal,Jafarzadeh}.

Future efforts at \bp\ tracking may include algorithm refinements.
\citet{Berger 1998} differ from our approach in their ability to sensibly track \bp s through merging and splitting events, whereas we simply consider the products of such events (which are fortunately rare in our \muram\ data set) to be newly-found \bp s.
Their algorithm is also resilient to \bp s which disappear for up to a few frames before reappearing.
However, this improvement may become unnecessary with the high cadence and sensitivity of DKIST data.
Algorithmic improvements to improve robustness or detection quality might also be adapted from the recommendations of \citet{Craig}, benefiting from existing efforts in large-scale feature detection.

\acknowledgements

The authors warmly thank Matthias Rempel for sharing the results of his \muram\ simulations, and Mark Rast and Piyush Agrawal for discussing with us the results of their own work and inspiring the contents of this paper's appendix.
SJV thanks Craig DeForest and Derek Lamb for a productive summer spent learning many coding techniques used in this analysis.
This work was supported by NSF grant 1613207 (AAG).
SRC's work was also supported by NASA grants {NNX\-15\-AW33G} and {NNX\-16\-AG87G}, NSF grant 1540094 (SHINE), and start-up funds from the Department of Astrophysical and Planetary Sciences at the University of Colorado Boulder.
Portions of this work utilized the RMACC Summit supercomputer, which is supported by the National Science Foundation (awards ACI-1532235 and ACI-1532236), the University of Colorado Boulder, and Colorado State University. The Summit supercomputer is a joint effort of the University of Colorado Boulder and Colorado State University.
This work made extensive use of the Scientific Python ecosystem and NASA's Astrophysics Data System.

\appendix

\section{Centroid Jitter}
\label{sec:jitter}

The tracking of centroids in identified features is susceptible to jitter in the centroid motion due to the effect of feature shape changes on the computed centroid location.
Whether the shape change is due to twisting or compression of the \bp , the merging of sub-resolution \bp s with the tracked \bp , or is just an artifact of the tracking routine due to changes in the identified edge of the \bp , a change in shape can produce additional centroid motion beyond idealized, rigid-body advection of the \bp .
This additional motion may represent real motion of magnetic field lines and a source of higher-order MHD waves.
However, in the context of tracking \bhp\ centroids as representative of the flux tube's bulk motion (the context of this and most other papers tracking \bp s as MHD wave sources), this additional motion is spurious.
Centroid tracking becomes vulnerable to this effect only when increased spatial resolution changes \bp s from points into well-defined features---i.e. high-resolution simulations and DKIST observations.
This effect is explored more fully in \citet{Agrawal}; here we attempt to constrain the influence of this effect on power spectra such as those presented in this paper.

Between two consecutive Frames 1 and 2, a \bp\ will have measured areas $A_1$ and $A_2$, with a size change $\Delta A \equiv A_2-A_1$.
In this section we use a simple model of the Frame 1 \bp\ as a square of side length $\sqrt{A_1}$ in Frame 1, and the \bp\ in Frame 2 as a rectangle of side lengths $\sqrt{A_1}$ and $\sqrt{A_1} + \ell$, where $\ell = \Delta A / \sqrt{A_1}$ so that the rectangle's area is $A_1 + \Delta A = A_2$.
This is an intermediate case for the effect of a size change $\Delta A$ on the centroid location, where the extremes are uniform, radial expansion having no effect on the centroid, and the addition of a long and thin filament to the shape of Frame 1 having an extreme effect on the centroid.
In this model, and assuming no advection of the \bp , the unweighted centroid will move a distance of $\Delta x = \Delta A / (2 \sqrt{A_1})$ along the major axis of the rectangle.

We calculated $\Delta x$ from measured areas for each \bp\ and for each timestep in our 20-second-cadence \muram\ simulation.
We did the same after applying our tracking routine to a 2-second-cadence \muram\ run (without the 30~G net injected flux of the 20-second-cadence run; see \citet{Rempel,Agrawal}).
While this 2-second-cadence run has a smaller spatial domain (6~$\times$~6~Mm, versus the 24.5~$\times$~24.5~km of the 20-second-cadence simulation used in this paper) and thus provides fewer \bp s and poorer statistics, it allows us to investigate how DKIST measurements might be affected by this centroid jitter. 

\begin{figure}[h]
	\centering
	\includegraphics[width=0.45\textwidth]{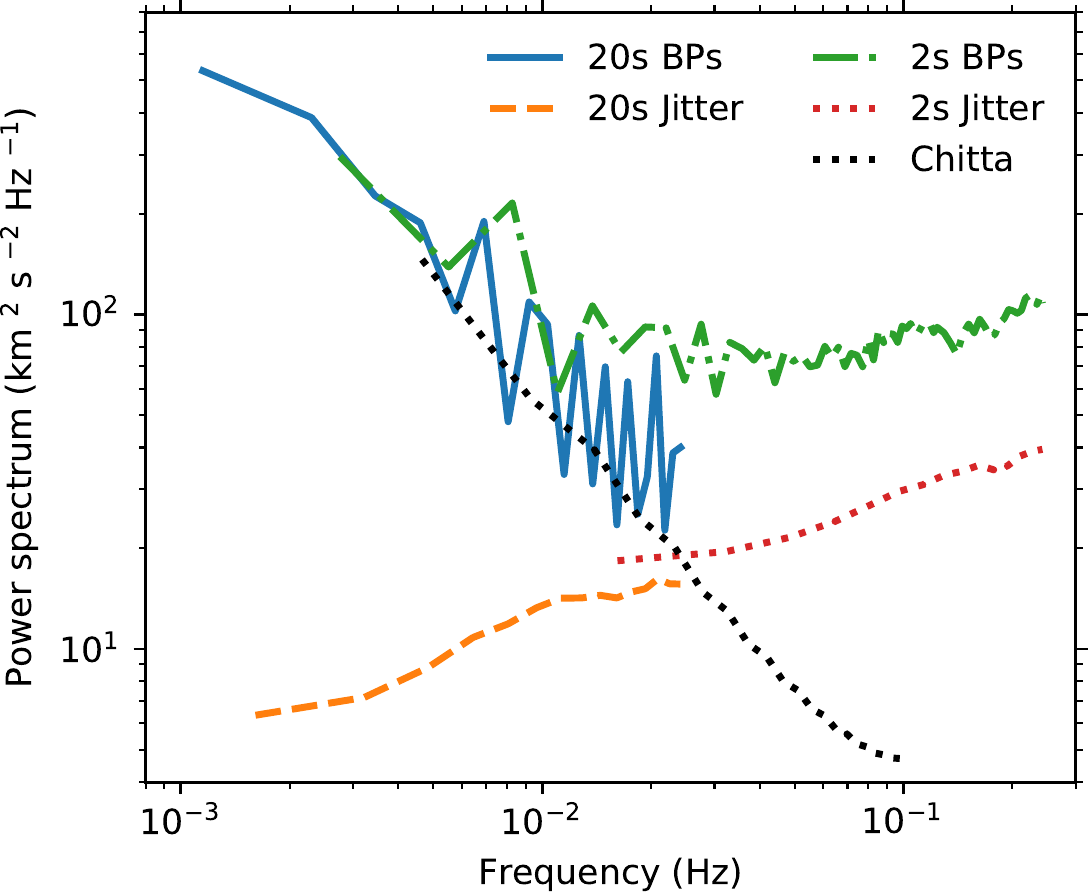}
	\caption{Power spectra of the total \bp\ (BP) motion and the modeled contribution to that motion of centroid jitter for \bp s in 20-second- and 2-second-cadence \muram\ runs. The 20-second-cadence \bp\ spectrum is the same as is shown in Figure \ref{fig:all_spectra}, and the observational \citet{Chitta} spectrum of that figure is also reproduced to facilitate comparison.}
	\label{fig:jitter}
\end{figure}

To simulate the contribution of centroid jitter to the motion of a \bp , we then average the power spectra of every 30-sample sequence of $\Delta x$ values.
These spectra are shown in Figure \ref{fig:jitter}.
The jitter spectra for the two different cadences produce a nearly-continuous line.
The spectrum for the 20-second-cadence data exceeds the jitter spectrum while showing a consistent slope of opposite sign.
The spectrum for the 2-second-cadence data, however, coincides with the 20-second-cadence spectrum for low frequencies, but levels off and even takes on a positive slope for the highest frequencies.
In the mid- to high-frequency regime, this spectrum suggestively takes the shape of the jitter spectrum.

We interpret this to mean that, at a 20-second-cadence, the jitter in a \bp 's centroid is small compared to the advective motion between frames, and the analysis in this paper is safe from misinterpreting jitter as bulk advection.
However, the additional frequencies probed by the 2-second-cadence data appear to be dominated by jitter rather than advection.
We base this on the similarity in slope between the \bhp\ and jitter spectra and attribute the factor of $\sim 2$ difference in magnitude to the simplicity of our jitter model.
This effect may not have been visible in the high-cadence \citet{Chitta} observations due to their spatial resolution rendering unresolvable the shape changes that give rise to this jitter.
However, we anticipate both high cadence and high spatial resolution in DKIST data.
Therefore, analysis of MHD wave excitation by \bp s observed by DKIST must take a more advanced approach or it will risk assigning the energy in higher-order modes to simple, bulk advection.

\end{document}